\documentclass[utf8]{frontiersSCNS} 

\usepackage{url,hyperref,lineno,microtype,subcaption}
\usepackage[onehalfspacing]{setspace}

\def\keyFont{\fontsize{8}{11}\helveticabold }
\def\firstAuthorLast{Paiano {et~al.}} 
\def\Authors{Simona Paiano\,$^{1,2}$, Renato Falomo\,$^{1}$, Marco Landoni\,$^{3}$, Aldo Treves\,$^{4}$ and Riccardo Scarpa\,$^{5,6}$}
\def\Address{$^{1}$INAF, Osservatorio Astronomico di Padova, Vicolo dell'Osservatorio 5 I-35122 Padova (PD) - ITALY \\
$^{2}$INFN, Sezione di Padova, via Marzolo 8, Padova - ITALY \\
$^{3}$INAF, Osservatorio Astronomico di Brera, Via E. Bianchi 46 I-23807 Merate (LC) - ITALY\\
$^{4}$Universit\`a degli Studi dell'Insubria, Via Valleggio 11 I-22100 Como - ITALY \\
$^{5}$Instituto de Astrofisica de Canarias, C/O Via Lactea, s/n E38205 - La Laguna (Tenerife) - ESPANA\\
$^{6}$Universidad de La Laguna, Dpto. Astrofsica, s/n E-38206 La Laguna (Tenerife) - SPAIN  }

\def\corrAuthor{Simona Paiano}
\def\corrEmail{simona.paiano@oapd.inaf.it}

\begin{document}
\onecolumn
\firstpage{1}

\title[Optical view of blazars]{An optical view of extragalactic $\gamma$-ray emitters} 

\author[\firstAuthorLast ]{\Authors} 
\address{} 
\correspondance{} 

\extraAuth{}

\maketitle

\begin{abstract}

\section{}
The Fermi Gamma-ray Observatory discovered about a thousand extragalactic sources emitting energy from 100 MeV to 100 GeV. 
The majority of these sources belong to the class of blazars characterized by a quasi-featureless optical spectrum (BL Lac Objects).
This hampers the determination of their redshift and therefore hinders the characterization of this class of objects.
To investigate the nature of these sources and to determine their redshift, we are carrying out an extensive campaign using the 10m Gran Telescopio Canarias to obtain high S/N ratio optical spectra. 
These observations allow us to confirm the blazar nature of the targets, to find new redshifts or to set stringent limits on the redshift based on the minimum equivalent width of specific absorption features that can be measured in the spectrum and are expected from their host galaxy, assuming it is a massive elliptical galaxy.
These results are of importance for the multi-frequencies emission models of the blazars, to test their extreme physics, to shed light on their cosmic evolution and abundance in the far Universe.
These gamma emitters are also of great importance for the characterization of the extragalactic background light through the absorption by the IR-optical background photons.

\tiny
 \keyFont{ \section{Keywords:} Blazars, BL Lac objects, optical spectroscopy, redshift, $\gamma$-ray astronomy} 
\end{abstract}

\section{INTRODUCTION}
A blazar is a jetted active galactic nucleus (AGN) with the relativistic jet that points along the line of sight of the observer. 
These kind of objects are bright emitters at all frequencies (from radio to TeV regime ), are characterized by high variability at all frequencies and large polarization, and are often dominated  by the $\gamma$-ray emission especially during the outbursts.

Their spectral energy distribution (SED) shows the typical doubled-humped structure with two broad peaks: the first bump is located at low energies, typically in the infrared to X-ray region, and is interpreted as due to synchrotron emission produced by electrons of the jet spiraling along the lines of force of the magnetic field, instead the second peak is placed at higher frequency, between the X-ray and the MeV-TeV energies, and as suggested in most leptonic models, can be due to Compton scattering of the same electrons \citep[e.g.][]{maraschi1992, dermer1993, ghisellini2009a}.

Blazars are commonly classified in two categories, BL Lac Objects (BLLs) and the Flat Spectrum Radio Quasars (FSRQs), and this classification depends on the strength of their broad emission lines respect to the continuum. 
A more physical distinction is based on to the comparison between the luminosity of the broad line region (BLR) and the Eddington luminosity  \citep[e.g.][]{ghisellini2017}: FSRQs have radially efficient accretion disk, while the BL Lac objects are not able to photoionise gas of the BLR, causing the absence of features in the majority of their optical spectra.
This classification needs to know the mass of the accreting black hole and the redshift of the source, which for broad emission line AGNs can be determined by spectroscopy, while for the BLLs is arduous due to the faintness or absence of the emission/absorption lines, showing a completely featureless optical spectra \citep{falomo2014}. 

The advent of the \textit{Fermi} $\gamma$-ray observatory, starting observations in 2008 \citep{atwood2009}, with its systematic scanning of the entire sky every 3 hours at the high energy band (HE; $>$ 20~MeV), has revolutionized the blazar studies, previously performed with radio and X-ray surveys. 
The extragalactic $\gamma$-ray sky is dominated by blazars \citep{stephens2015} and in the third AGN \textit{Fermi}/LAT catalogue \citep[3LAC]{ackermann2015}, 1738 blazars are reported, compared with the 3033 $\gamma$-ray  detected sources:662 are classified as BLL, 491 as FSRQ, while the remaining blazars are as of uncertain type.
It is worth to note that a growing sub-sample of the GeV blazars are also emitters at the TeV energies (VHE; E$>$100 GeV), detected by the Cherenkov telescopes as MAGIC, VERITAS and the HESS arrays, that can sample energies down to 30-50 GeV. The majority of them are BLLs (in the TeVcat\footnote{http://tevcat.uchicago.edu/} there are 60 BLLs against 6 FSRQs), implying that the BLL class dominates the extragalactic TeV sky.

Although BLLs are the most numerous extragalactic class in the HE and VHE bands, for a large fraction of them the redshift is still unknown or highly uncertain, because contrarily to most AGNs, the BLLs are characterized by featureless (or quasi) featureless spectra.
On the basis of the recent statistics, it was proposed that on average BLLs have lower redshift and smaller high energy  $\gamma$-ray luminosity than FSRQs \citep{ghisellini2017}.
However, this could be due to a bias since the number of robustly detected high redshift BLLs is significantly limited due to the difficulty to measure their distance. 
Hence the determination of the blazar redshift is crucial to calculate their luminosity, to build and characterize realistic emission models and to allow us a sound comparison of the multi-frequency SEDs between the two blazar classes (see the \textit{blazar sequence}, \citet{fossati1998} and \citet{ghisellini2017}).

The estimation of the BLL redshifts is also essential to determine the properties of the extragalactic background light (EBL).  The BLL $\gamma$-rays of high energy can interact with the EBL infrared-optical photons and produce pairs $e^{-}$/$e^{+}$, resulting in an absorption in the GeV-TeV BLL spectrum starting at frequencies and with optical depth that depend on the redshift of the $\gamma$-ray source and is more pronounced in the 0.5 $<$ z $<$ 2 interval \citep{franceschini2008}.
At higher $z$ the absorption due to pair production moves to \textit{Fermi} energies, completely extinguishing the source in the VHE regime.
Although a significant number of FSRQ detections, up to z $>$ 4 already exist \citep{ackermann2017}, at the TeV energies, due to their Compton inverse peak position, only a small fraction of them are detected. 
Therefore the identification of high redshift BLLs at these energies is particularly challenging in order to study the earliest EBL components due to the first-light sources (Population III stars, galaxies or quasars) in the universe \citep[e.g.][]{franceschini2017}.

These considerations motivated us to carry out an extensive campaign at the 10.4 m Gran Telescopio Canarias aimed to obtain high signal-to-noise ratio (SNR) optical spectra and to estimate the redshift of BLLs. 
The results of this project are shown in \citet{landoni2015} for S20954+65, \citet{paiano2016} for the TeV BLL S20109+22, \citet{falomo2017} for the blazar B0218+357, \citet{paiano2017tev} for a sample of 15 TeV BLL and 7 TeV candidates with unknown or uncertain redshift and in \citet{paiano2017fgl} focused on 10 BLLs detected by \textit{Fermi} satellite suggested to be at relatively high redshift by previous works. 
Moreover, till now, we observed 40 unassociated $\gamma$-ray objects detected by \textit{Fermi} and candidate to be blazars (Paiano et al., 2017, submitted ), 20 BLLs optically selected among the SDSS blazars (Landoni et al., 2017, submitted) and 10 hard \textit{Fermi} sources.

\section{SAMPLE, DATA REDUCTION AND ANALYSIS}

This spectroscopic program involved a conspicuous sample of BLLs, for a total of $\sim$ 100 objects and for which the selection followed different criteria \citep[see details in ][]{paiano2017tev, paiano2017fgl}. 
All of these objects were also selected to have unknown or uncertain redshift for which conflicting estimates are published in the literature, mainly due to spectra with low S/N. 

The observations were gathered at the GTC using the medium resolution spectrograph OSIRIS \citep{cepa2003}, configured with the grisms R1000B and R1000R and covering the spectral range 4100-10000 $\textrm{\AA}$.
Details about the observational strategy, the data reduction and analysis procedure are reported in \citet{paiano2017tev} and \citet{paiano2017fgl}, where all spectra, corrected for atmospheric extinction and flux calibrated, are reported. 
They are also reported in the website http://www.oapd.inaf.it/zbllac/.
On average, the S/N ranges from 100-200 at 4500  $\textrm{\AA}$ and 8000 $\textrm{\AA}$, to a peak of 300 at $\sim$6000 $\textrm{\AA}$.

\section{RESULTS}
Our spectra allow us to confirm the blazar nature of the observed targets and, on the basis on the spectroscopic properties, they exhibit four characteristic features: 1) weak emission lines typical of low density gas; 2) spectral lines of stars from the host galaxy; 3) intervening absorption systems due to cold gas; and 4) a pure featureless continuum.
While the first three types can co-exist and from them a redshift can be derived when emission and/or absorption lines are identified, in the latter case, thank to high SNR achieved, we can set a stringent redshift lower limit on the basis of the minimum Equivalent Width (EW) of absorption lines expected from the starlight emission of the blazar host galaxy.

\begin{itemize}
\item \textbf{Spectra with emission lines characteristic of low-density gas}

Although the BLL optical spectra are mainly characterized by a featureless continuum, very weak emission lines can be seen, with an intrinsic luminosity lower than those observed in quasars. 
Owing to the weakness of these lines, their detectability depends on the state of the source, with faint state of the source that favors the detection of intrinsic emission lines, and on the quality of the observations, especially in terms of S/N ratio.

In particular from our spectra, we detect weak emission lines due to [O III] (5007 $\textrm{\AA}$) in the spectrum of 1ES1215+303 , W-Comae, MS1221.8+2452 and PKS1424+240, and the [O II] (3727 $\textrm{\AA}$) emission in 1ES0033+595, 1ES1215+303 and PKS1424+240.
For the first time, we are able to determine the spectroscopic redshift of z~=~0.467 and z~=~0.6047  for 1ES0033+595 and PKS1424+240 respectively (in Fig. 1 the spectrum of PKS1424+240, one of the farthest TeV BLLs).

\begin{figure}[h!]
\begin{center}
\includegraphics[width=7cm, angle=-90]{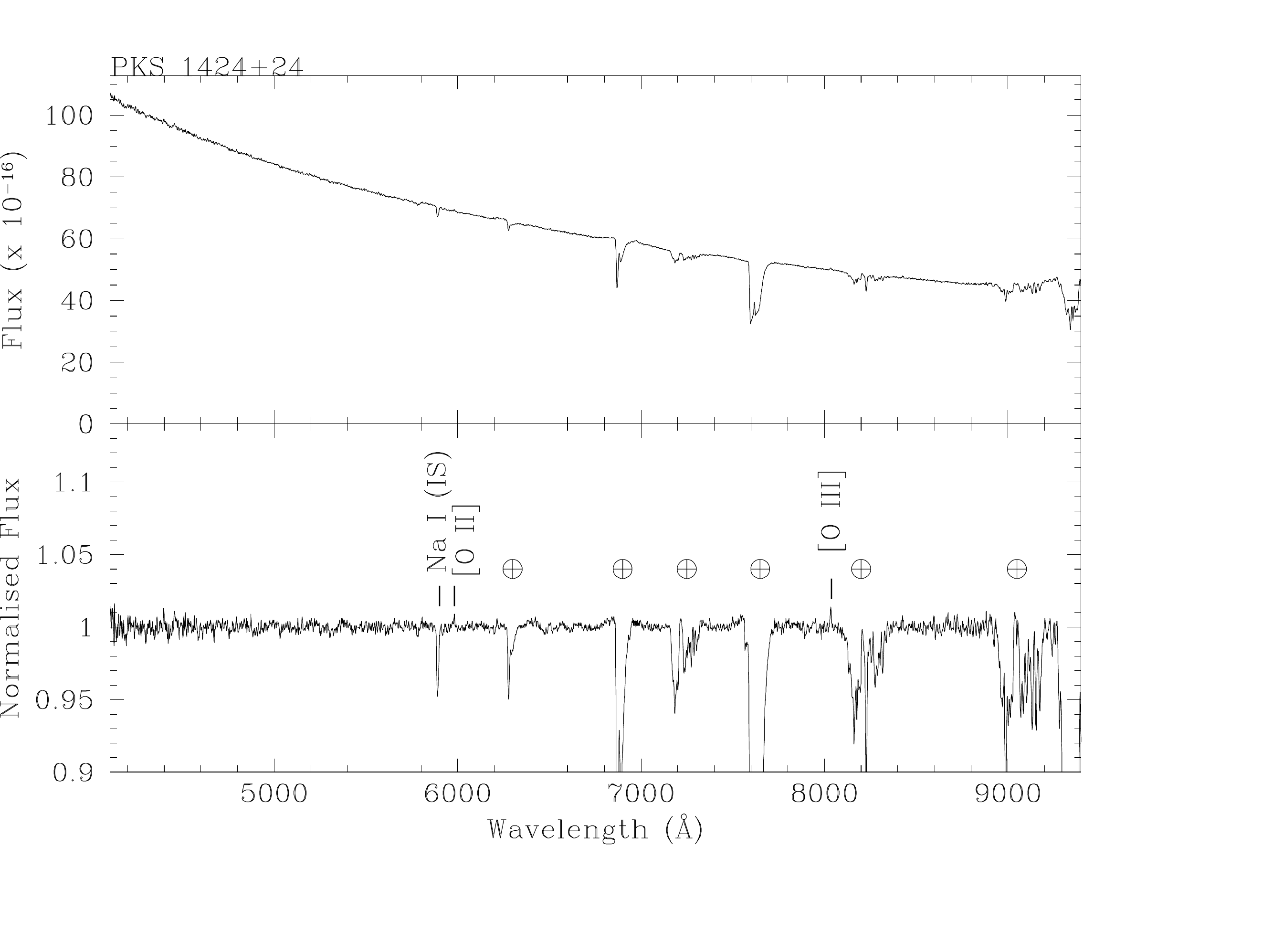}
\end{center}
\begin{minipage}[c]{1.\textwidth}
 \begin{center}
    \includegraphics[width=5cm, angle=-90]{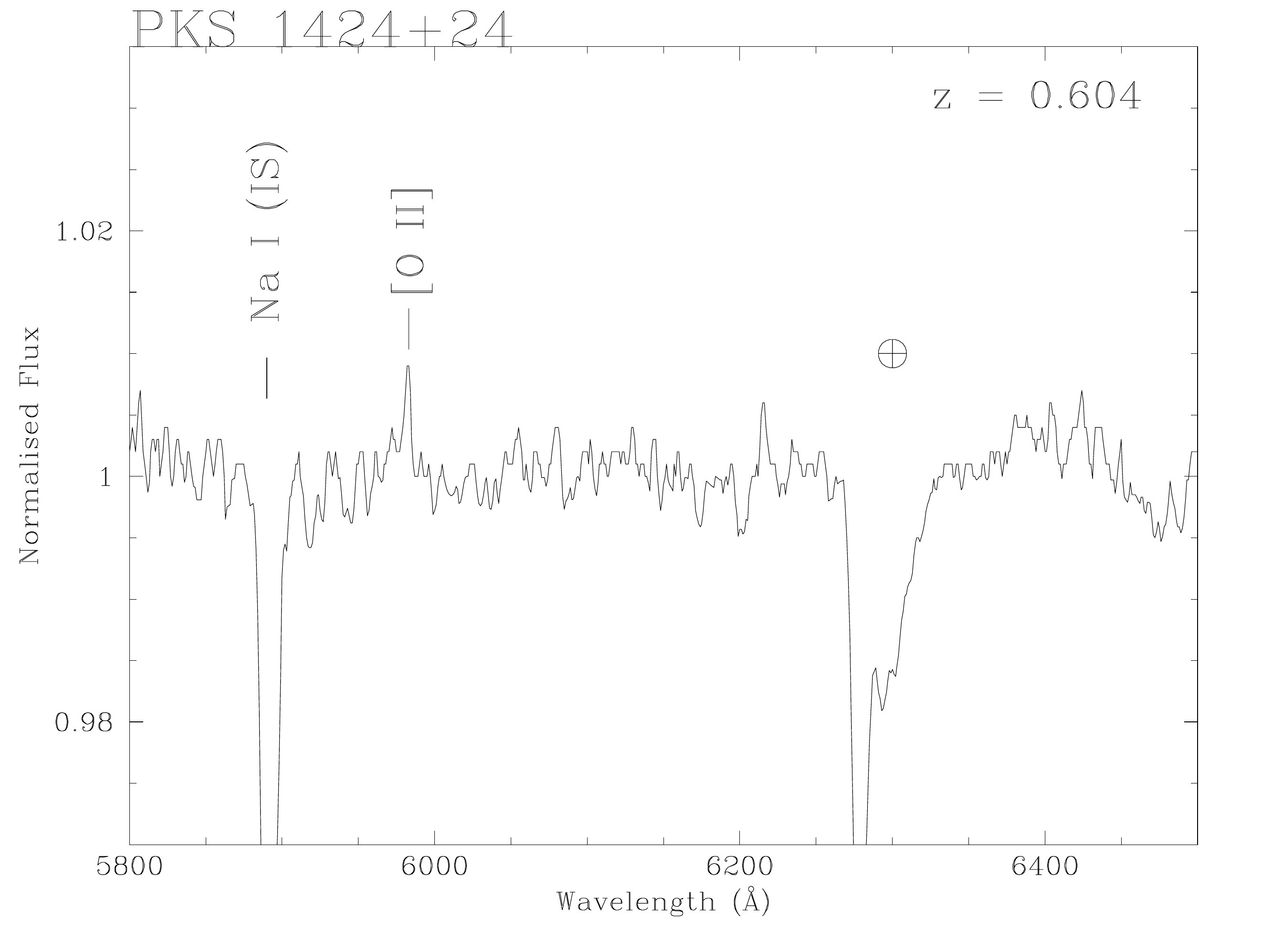}
    \includegraphics[width=5cm, angle=-90]{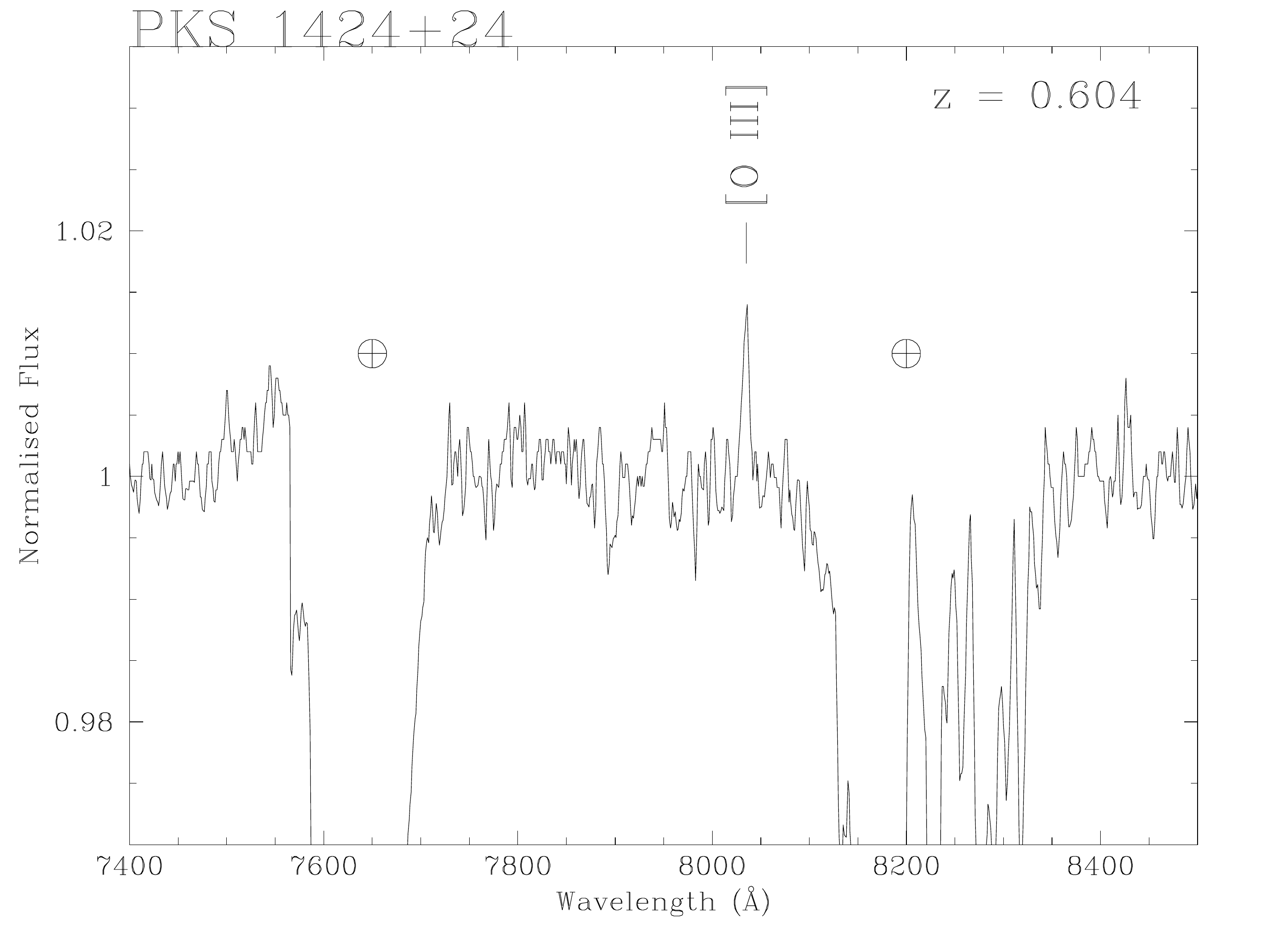}
    \end{center}
    \end{minipage}
\caption{Spectrum of the TeV source PKS~1424+240 (z~=~0.6047) obtained at GTC. \textit{Top panel}: Flux-calibrated and dereddered spectrum, and normalized spectrum. The main telluric bands are indicated by $\oplus$, the absorption features from interstellar medium of our galaxies are labelled as IS (Inter-Stellar). \textit{Bottom panels}: Close-up of the normalized spectrum around the found emission spectral lines, marked by line identification.} \label{fig:1}
\end{figure}

\item \textbf{Spectra with absorption lines due to the stellar population of the host galaxy}

BLLs are located in the nuclei of giant/massive elliptical galaxies with a prominent spheroidal component and with average luminosity in the R band of M$_R \sim$~23 \citep[][and references therein]{falomo2014}. 
Their stellar population is composed by old stars and therefore the main observable absorption lines are Ca~II (3934,3967 $\textrm{\AA}$) , G-band (4304 $\textrm{\AA}$), H$_{\beta}$ (4861$\textrm{\AA}$), Mg~I (5175 $\textrm{\AA}$) and Na~I (5875 $\textrm{\AA}$). 
These absorption features can be detected over the non-thermal component and this depends on the signal-to-noise ratio and the spectral resolution of the spectrum, and finally on the relative power of the non-thermal nucleus and the host galaxy.
In very few cases, high quality spectroscopy with adequate high spatial resolution can probe star formation of the host galaxy and allow us to detect narrow emission lines from it. 

In our sample, we were able to detect host galaxy absorption lines for W Comae (z~=~0.102), MS 1221.8+2452 (z~=~0.218), 3FGL J0505.5+0416 (z~=~0.423) and 3FGL J0814.5+2943 (z~=~0.703, see Fig. 2).

\begin{figure}[h!]
\includegraphics[width=7cm, angle=-90]{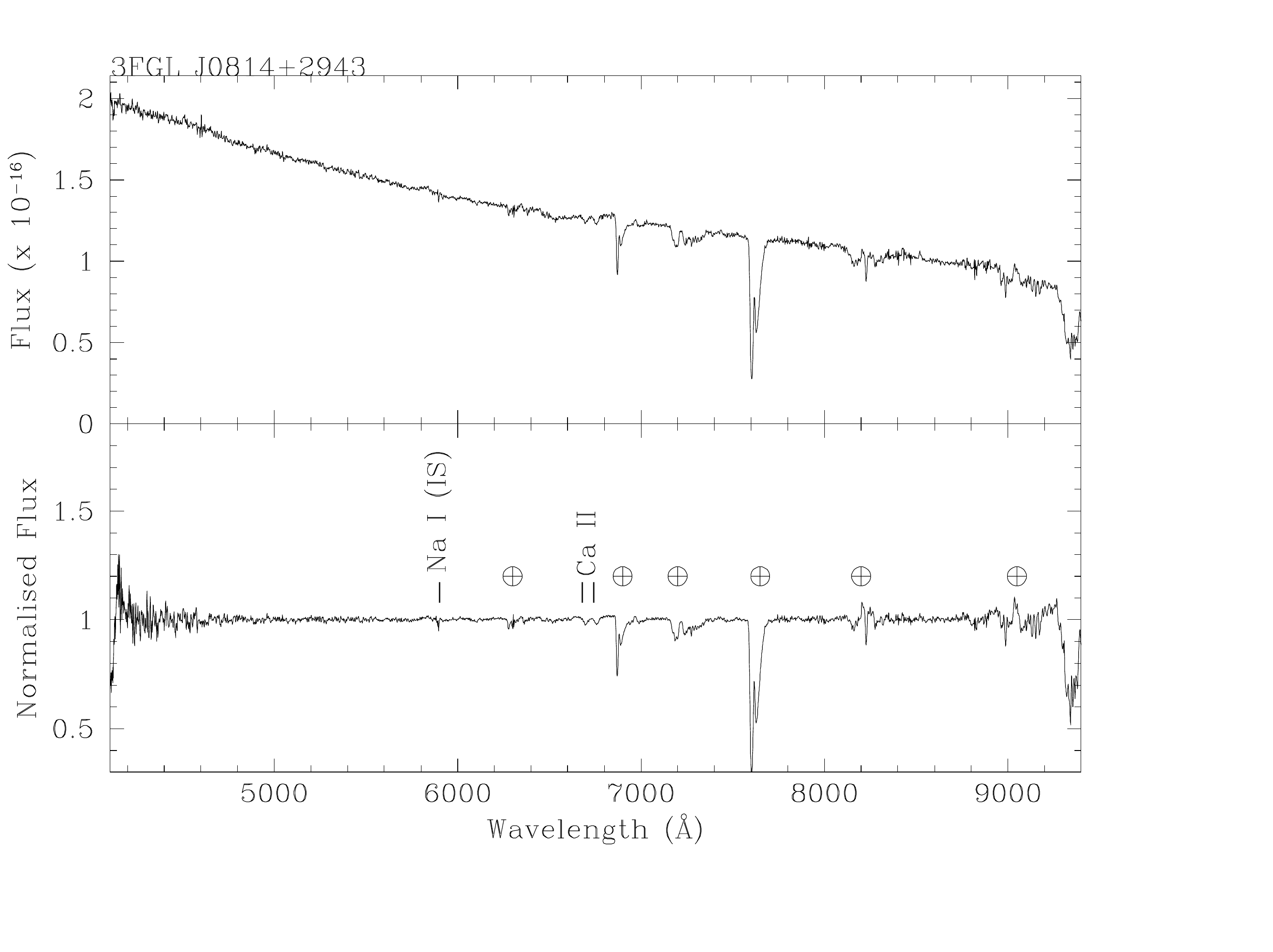}
\includegraphics[width=6.3cm, angle=-90]{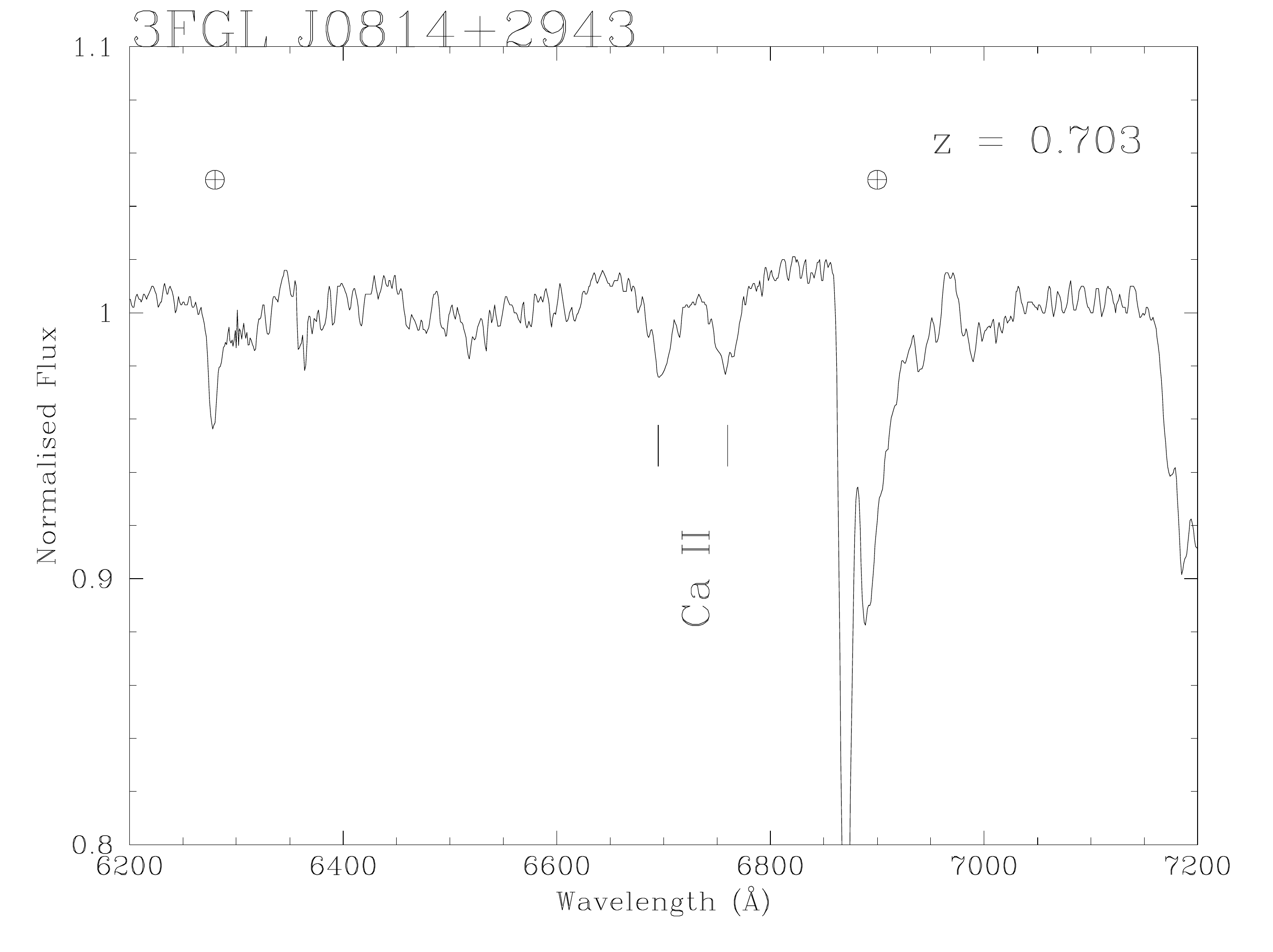}
\caption{Spectrum with S/N~$\sim$~160 of the 3FGL BLL J0814.5+2943. In the close-up image, the absorption doublet at 6699-6759 $\textrm{\AA}$ identified as Ca II (3934,3967 $\textrm{\AA}$) at z~=~0.703 from the the host galaxy.}\label{fig:2}
\end{figure}

\item \textbf{Spectra with intervening absorption lines}

As for the high redshift quasars, gas along the BLL direction can produce systems of absorption at redshifts lower than the target redshift. Given that BLLs show quasi featureless continuous spectrum, they are ideal sources for studying absorptions from these intervening systems \citep[][and references therein]{landoni2012}.  
The redshift of the intervening systems yields a lower limit to the BLL redshift, while an upper limit can be set from the expected distribution of the absorbers \citep{zhu2013}. 

For six of our targets,  3FGL J0008.0+4713, BZB J1243+3627, and 3FGL J1450.9+ 5200 with an uncertain redshift in literature, and BZB J1540+8155, 3FGL J1107.5+0222, and BZB J2323+4210 with unknown redshift, we detected intervening absorption systems that allow us to set spectroscopic redshift lower limits.
In Fig. 3 two examples of spectra of our sample that exhibit Ly$_{\alpha}$ forest, C IV (1548 $\textrm{\AA}$) and Mg~II (2800 $\textrm{\AA}$) absorption systems.

\begin{figure}[h!]
\includegraphics[width=7cm, angle=-90]{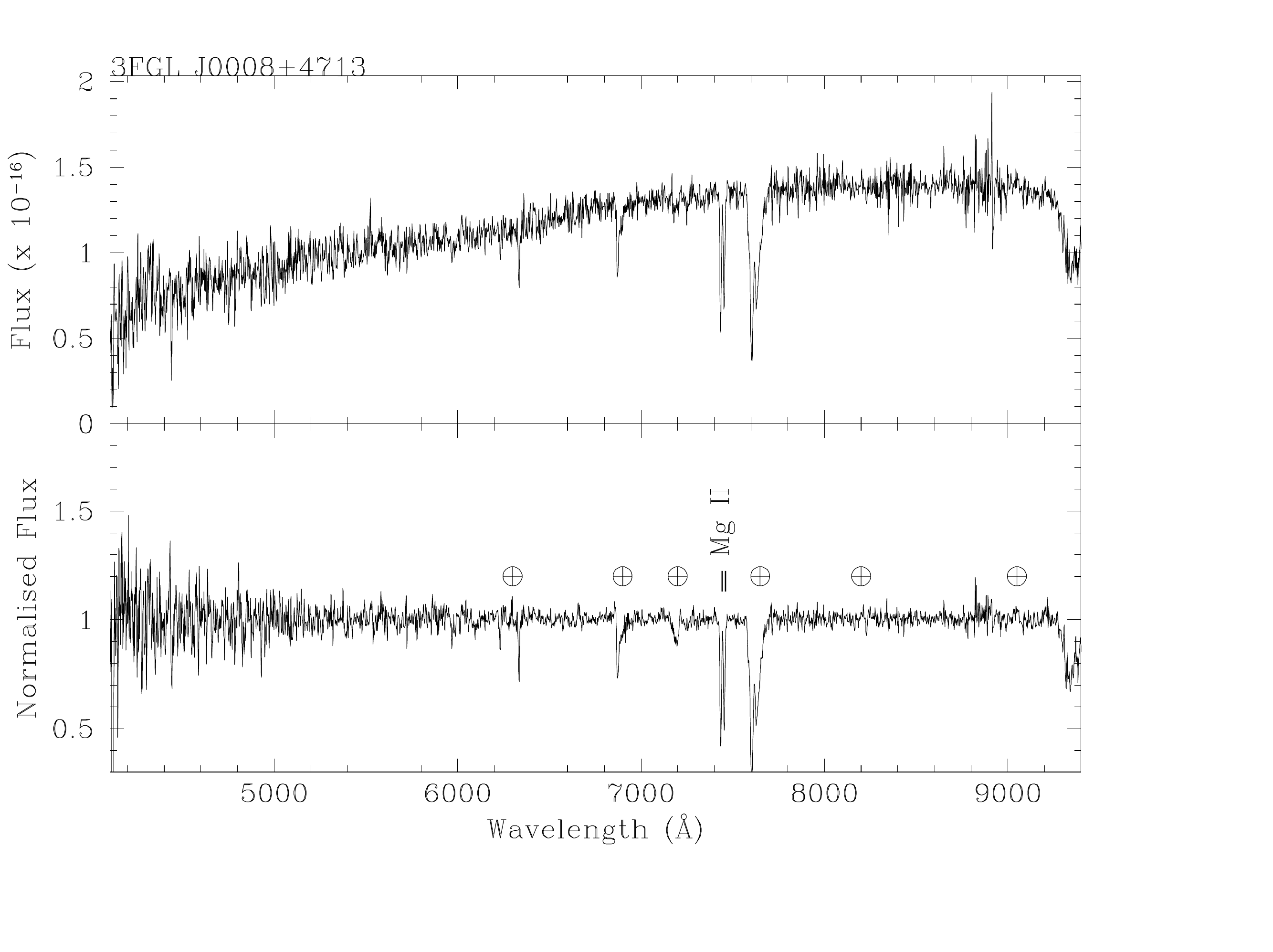}
\includegraphics[width=7cm, angle=-90]{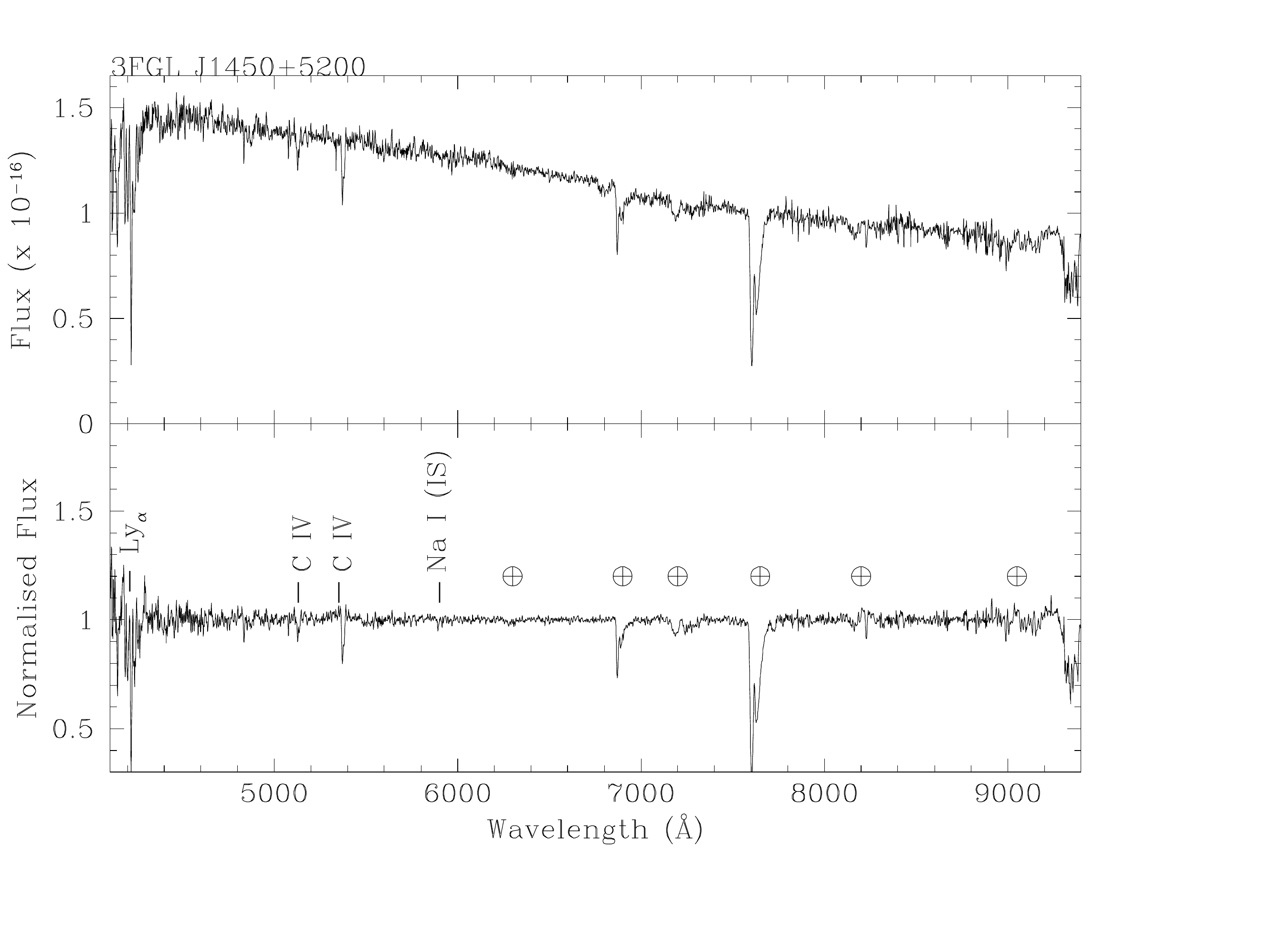}
\caption{Spectrum of the 3FGL BLLs J0008.0+4713, that shows an intervening doublet absorption system at 7440 $\textrm{\AA}$ identified as Mg II (2800 $\textrm{\AA}$) at z = 1.659, and the spectrum of the 3FGL J1450.9+ 5200 that shows the absorption line at 5372 $\textrm{\AA}$ due to C IV (1548 $\textrm{\AA}$) intervening gas and a strong Ly$_{\alpha}$ (1216 $\textrm{\AA}$) absorption at the same redshift (z~=~2.470). A second C IV intervening system at z = 2.312 is associated to the absorption feature detected at 5127 $\textrm{\AA}$.}\label{fig:3}
\end{figure}

\item \textbf{Featureless spectra}

In spite of the high quality of the optical spectra for 18 observed targets the spectrum appears very featureless.
This occurs when the emission from the underlying nebulosity of the host galaxy is over-shined by the non-thermal continuum.

As example, Fig. 4 shows the featureless spectra of two TeV BLLs RGB~J0136+391 and 3C~66A.
For the latter, a previous redshift of 0.444 was proposed in literature \citep{miller1978} and reported in the NED.
On the basis of our high S/N ratio and featureless spectrum, we cannot support this value and thus the redshift of 3C~66A is still unknown and the emission models of the source assuming the wrong redshift should be revised.

In the case of no detection of spectral lines, following the scheme outlined in Appendix A of \citet{paiano2017tev}, we can set redshift lower limit using the minimum Equivalent Width method, based on the assumption that the BL Lac objects are hosted in massive elliptical galaxies and that hence the detection of their stellar absorption features depends on the SNR of their optical spectra , on the spectral resolution, and on the flux ratio between the nucleus and the host galaxy.
The redshift lower limit found for 18 objects with featureless spectra are summarized in Table 3 of \citet{paiano2017tev} and \citet{paiano2017fgl}.

\begin{figure}[h!]
\includegraphics[width=7cm, angle=-90]{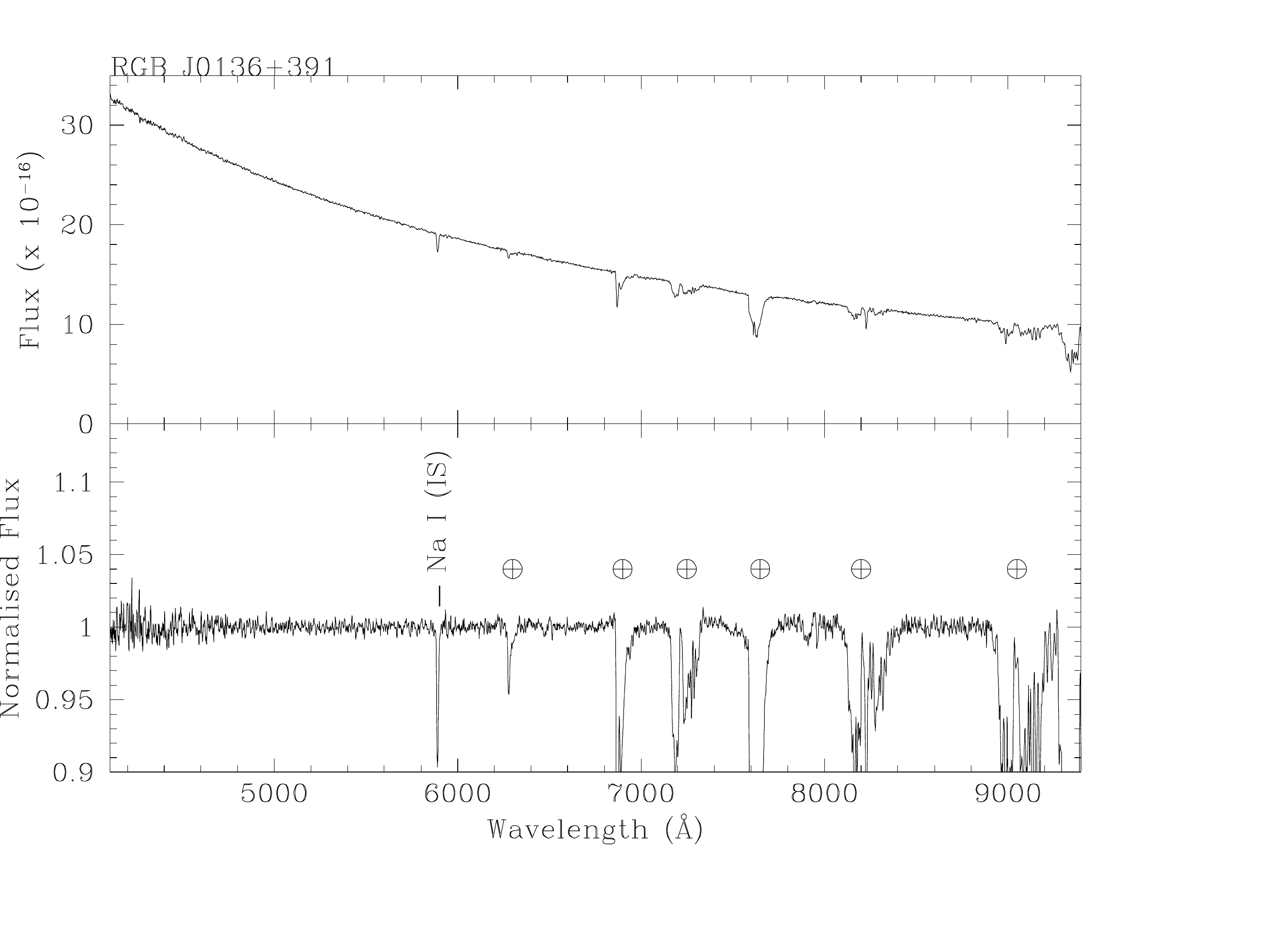}
\includegraphics[width=7cm, angle=-90]{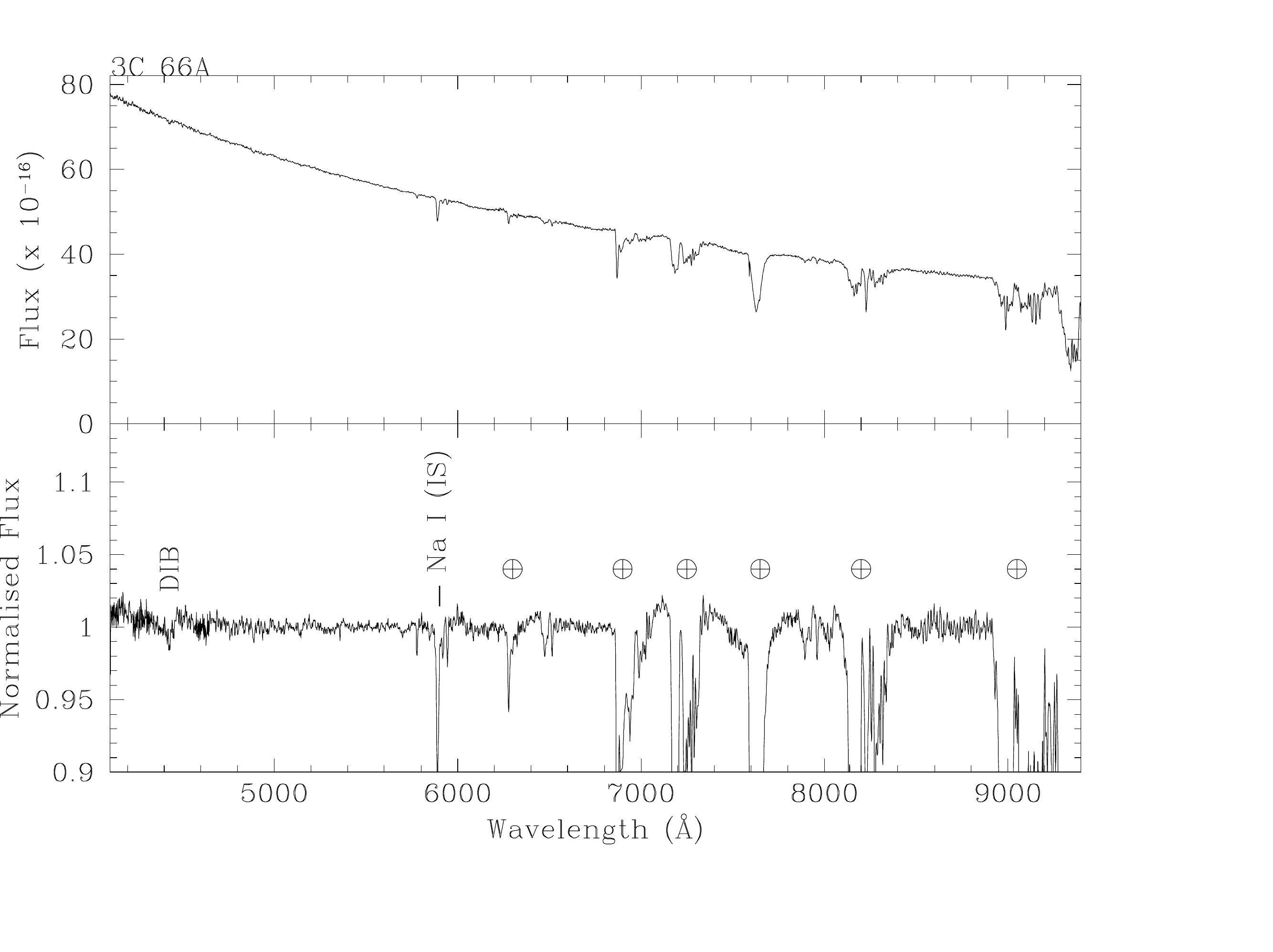}
\caption{ Featureless spectra of the TeV BLL RGB J0136+391 and 3C 66A. No emission/absorption lines are identified, hence for these sources the redshift is still unknown.}\label{fig:4}
\end{figure}

\end{itemize}

\section{Conclusion}

We secured high quality and S/N spectra for 32 $\gamma$-ray BL Lac objects, selected for being TeV sources, TeV candidates or high redshift \textit{Fermi} objects, with unknown redshift or with conflicting redshift values in literature.
We determined the redshift for 8 objects and spectroscopic redshift lower limit for 6 sources, including three of the farthest BLLs detected in the GeV and TeV regime, but for the remaining sources, in spite of the very high S/N and of the improvement of spectral resolution, their spectrum is featureless.

Although these observations represent the state of the art of the observing facilities (telescopes with large aperture and modern instrumentation) and the capabilities to perform the spectroscopy of BLLs, the redshift determination of this class of objects remains still rather arduous. 
For this reason, the detection of the spectral features in their spectra requires observations with very high SNR and an adequate resolution that seems feasible only with the next generation of extremely large (30 - 50 m class) telescopes, such as E-ELT (the European Extremely Large Telescope).
An idea about the performances of these new class of telescopes is given in \citet{landoni2014} where simulations of BLL spectra are made using an X-Shooter-like instrument (with an spectral resolution $\sim$ 3000) coupled to the E-ELT and including a moderate adaptive optics module.
With these new facilities, it should be possible to measure the redshift of BLLs having extreme nucleus-to-host galaxy ratios between $\sim$300 to $\sim$2500, depending on the redshift and the target brightness.

\section*{Acknowledgments}

The authors of this work thank the organizers of the ``Quasars at All Cosmic Epoch'' conference.

\bibliographystyle{frontiersinSCNS_ENG_HUMS} 
\bibliography{paiano_biblio}




\end{document}